# Comparative Evaluation of Text and Audio Simplification: A Methodological Replication Study

**Prosanta Barai**
Eller College of Management
University of Arizona
*prosantabarai@arizona.edu*

**Gondy Leroy**
Eller College of Management
University of Arizona
*gondyleroy@arizona.edu*

**Arif Ahmed**
Eller College of Management
University of Arizona
*arifahmed@arizona.edu*

**Abstract:**

This study serves as a methodological replication of Leroy et al.'s (2022) research, which investigated the impact of text simplification on healthcare information comprehension in the evolving multimedia landscape. Building upon the original study's insights, our replication study evaluates audio content, recognizing its increasing importance in disseminating healthcare information in the digital age. Specifically, we explored the influence of text simplification on perceived and actual difficulty when users engage with audio content automatically generated from that text. Our replication involved 44 participants for whom we assessed their comprehension of healthcare information presented as audio created using Leroy et al.'s (2022) original and simplified texts. The findings from our study highlight the effectiveness of text simplification in enhancing perceived understandability and actual comprehension, aligning with the original study's results. Additionally, we examined the role of education level and language proficiency, shedding light on their potential impact on healthcare information access and understanding. This research underscores the practical value of text simplification tools in promoting health literacy. It suggests the need for tailored communication strategies to reach diverse audiences effectively in the healthcare domain.

# 1 Introduction

Effective communication using clear and understandable language is essential in healthcare to promote health literacy. In the United States, improving health literacy is a national goal (Rep. Rangel et al., 2010; Baur 2010), as poor health literacy can lead to miscommunication, poor decision-making, and increased healthcare costs for patients (Eichler et al., 2009).

## 1.1 Original Study

In their original study, Leroy et al. (2022) evaluated a proprietary text simplification tool, which integrates algorithms (each known to affect text difficulty) to suggest concrete simplifications. After reading the original and simplified texts, the authors evaluated the tool's effectiveness by analyzing participants' performance using various metrics. This task arises due to the widespread use of text for information dissemination, which underscores the need to ensure that people can better understand, remember, and act upon healthcare information.

The original study's results indicated that simplified texts were perceived as less difficult, simplification led to improved performance on answering multiple-choice questions, and demonstrated better free recall of information based on automated metrics, suggesting the potential value of such tools for various audiences. Further analysis revealed correlations between self-reported education level and language spoken at home with question accuracy for original texts, which diminished with simplified texts.

## 1.2 Replication Study

We replicated the study to assess whether the improvements in health literacy observed with the simplified text also hold when the information is presented in an audio format generated from the tool's simplified text. This would validate whether the effectiveness of the simplification tool and its underlying principles extend to audio-based healthcare information delivery.

### 1.2.1 Text to Audio Scripts

According to Lim & Benbasat (2000), among the various channels of information delivery (e.g., print, audio, and video), text-based and multimedia representations are equally effective in increasing perceived understanding levels when it comes to analyzable tasks[1] (tasks where explicit facts are available). However, only multimedia representation effectively reduced perceived equivocality levels for less analyzable tasks[2] (tasks that require human judgment). According to Nielsen Norman Group, the written script plays a vital role in developing comprehensible video and audio materials (Schade, 2014). They emphasized its significance in video production, stating that the script is the fundamental basis for the entire project. This highlights the connection between a well-constructed written script and creating simplified and understandable audio content.

### 1.2.2 The Need for Audio Content Simplification

The shift towards audio-based information delivery is significant. Recent data highlights the growing adoption of voice technologies: in 2019, smart speakers saw 19 million units sold, with 16% of health-related queries (Kinsella et al., 2021; Vailshery, 2019). Projections suggest smart speaker sales could reach 409.4 million by 2025, indicating a substantial market transformation (Vailshery, 2019), while audio-based technologies are increasingly favored by older adults (Kowalski et al., 2019; Wulf et al., 2014).

Despite the growing use of smart speakers and audio-based technologies for delivering medical information, the simplification of healthcare audio content has not been addressed. It can potentially enhance comprehension and patient compliance with medical instructions, particularly for populations with limited health literacy and impairments (Williams et al., 1998; Blair et al., 2019).

Therefore, by replicating the original study's methodology in an audio context, we seek to validate the universal applicability of text simplification tools and contribute to more effective, accessible healthcare communication strategies. We specifically sought to investigate whether the comprehension benefits observed in the original text-based study would persist when healthcare information is presented through automatically generated audio formats.

---

[1] In analyzable tasks, we have "predetermined responses to potential problems, and well-known procedures, are available and useful" Rice et al. (1992).

[2] In less-analyzable tasks, there aren't already known answers, procedures, or explicit knowledge about what is required to solve or complete the problem; instead, human judgment is required (Daft & Macintosh, 1981).

## 2  Materials and Method

### 2.1  Text Simplification

For this replication study, we reused the four texts (two long and two short) collected and simplified by Leroy et al. (2022) using their simplification tool. The shorter texts addressed pemphigus and polycythemia vera, while the longer texts focused on asthma and liver cirrhosis (see Appendix B). These texts covered common and less familiar medical topics, selected from Wikipedia, to ensure readers had low prior knowledge.

In the original study, a local health educator simplified the texts using the author's online editor. The educator received a brief introduction to the editor's functions and was instructed to utilize only the suggestions provided by the tool for text simplification. Importantly, no other constraints were imposed to maintain the study's external validity. It's worth noting that the questions evaluating content comprehension were not simplified, and the contained language was absent in the simplified and original texts.

The text simplification tool encompasses several valuable features, including lexical simplification, negation detection, grammar feedback, topic visualization, and audio generation (see Figure 1). Users input their text into a primary text box, and upon selecting the "Simplify Text" button, the tool offers algorithmic feedback by highlighting challenging words in blue and suggesting simpler alternatives. It also identifies complex sentence structures in red and guides simplification. Users have the flexibility to customize the tool's settings, including the level of simplification and variety. The newest text editor also offers generative language model (ChatGPT) based simplification. However, models like Llama-3 and GPT-3, trained on web and book data, can offer helpful examples, but their simplifications don't always follow established rules and standards. Additionally, these models can have adversarial effects like memorizing specific data (Dong et al., 2024) and producing incorrect or made-up information, known as hallucinations (Bai et al., 2024).

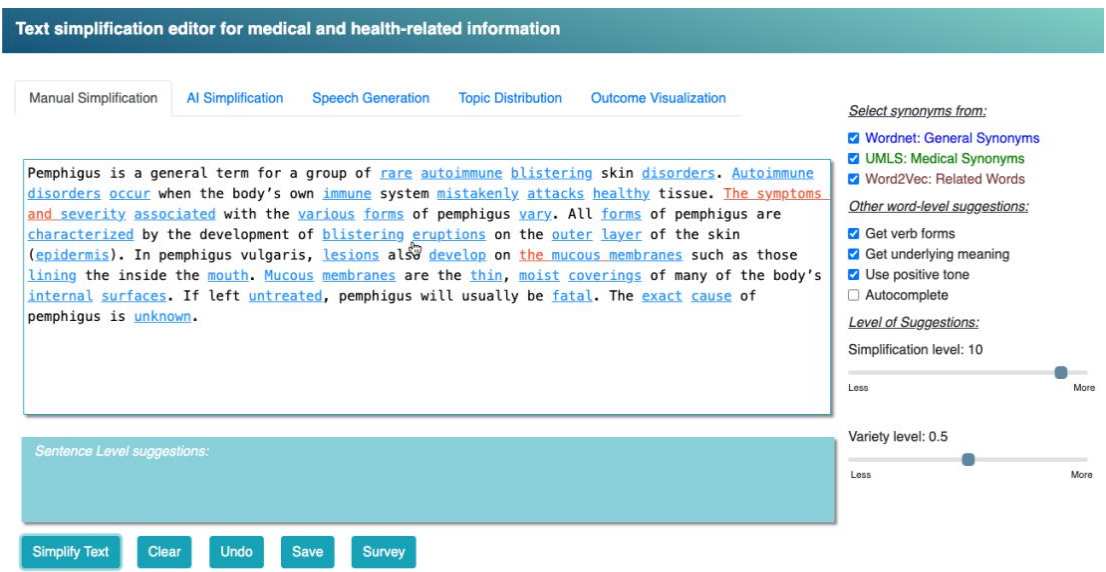

**Figure 1: Editor screenshot for the word-level suggestion.**

### 2.2  Audio Generation

We used Microsoft Azure Cognitive Service (speech engine 8) to generate audio using the REST API to convert text to speech. The Microsoft Azure Cognitive Services text-to-audio engine provides several tuning options including adjusting the speaking rate, volume, pitch, and voice style, as well as more advanced options, such as controlling the pauses between words and phrases and adjusting the pronunciation of specific words to enhance the audio quality further.

We generated audio files using a bit rate of 48khz, the highest available option, to ensure high-quality audio output, and we used an American English male voice. We chose the "Chat" style for the selected voice because it was more natural sounding among all the available options. We combined literature insights and cognitive judgment to configure the speech engine parameters. For example, Ahmed et al. (2024) found that appropriately placed pauses and emphasis can enhance information retention and reduce perceived difficulty. Based on this, we set the silence to "Landing Exact" with a 1000ms pause between sentences. Additionally, we adjusted the speech rate to 0.95 (default = 1), resulting in a moderate rate of 145 words per minute, which is considered optimal for improving comprehension (Ahmed et al., 2024). We used the same audio settings for all conditions of text difficulty.

## 2.3 Study Design

Our study adopted a randomized experimental crossover design, following the approach of the original study. Participants were randomly assigned to different experimental orders/ combinations (see Table 1), ensuring all possible unique research combinations were represented. This design choice allowed us to measure both the perceived and actual difficulty of the original and simplified content comprehensively (Kraemer et al., 1979).

The study comprised one independent variable: text difficulty (original vs. simplified). We examined two dependent variables: perceived difficulty and actual difficulty. Perceived difficulty refers to the participants' subjective assessment of how challenging they found the audio content upon initial encounter. In contrast, actual difficulty represents the content's objective impact on participants' understanding and comprehension, as measured by their performance on true-false, multiple-choice, and open-ended questions after engaging with the material.

We replicated the six combinations of the four experimental conditions to ensure a balanced distribution of order and audio topics. Data collection was conducted using REDCap (Research Electronic Data Capture) (Boateng et al., 2018), a secure and widely used platform in healthcare research. REDCap's randomization features for item presentation and answer options helped mitigate order effects and testing threats common in pre-post designs by reducing the risk of participants remembering and repeating answers across different conditions (Bhattacherjee, 2012). Additionally, its secure data management ensured the confidentiality and integrity of sensitive healthcare information.

Four audios related to asthma, polycythemia vera, pemphigus, and cirrhosis were presented in original and simplified versions to each participant, ensuring each participant participated in each condition without repeating the same topic. The participants were randomly assigned to one of six combinations. Each combination has two long and two short texts to balance the cognitive load and reduce bias, ensuring an accurate assessment of text simplification's impact. For example, Combination 1 includes Liver Cirrhosis, Pemphigus, Asthma, and Polycythemia Vera in that order, while Combination 2 includes Asthma, Pemphigus, Liver Cirrhosis, and Polycythemia Vera in a different order. Figure 2 shows a detailed breakdown of each combination according to the disease type of text used and the instrument used to collect observations. We first presented five true/false (TF) questions to assess participants' baseline knowledge of medical topics. Participants then listened to either the simplified or original audio, based on the combinations in Figure 2. Following the audio, they answered one general question (used as an attention check question), the five true/false questions, and five multiple-choice questions (MCQs). After completing these sections, demographic questions were administered. The arrows in Figure 2 illustrate the sequence of item presentation. The length of each audio ranges from 1.19 to 4.35 minutes.

We published the experimental conditions on Amazon Mechanical Turk (AMT), inviting participants to complete one of six experimental conditions. Each combination was assigned to 10 participants. Participants who agreed to participate were redirected from AMT to REDCap to complete the task. Upon completion, they were given a unique completion code to submit on AMT to enable us to track their response and approve or reject their submission associated with their AMT participant ID.

We implemented several additional properties to ensure the quality and timely completion of the task. For instance, each Human Intelligence Task (HIT) on AMT offered a compensation of $7.50 to participants who completed the assigned combination within sixty minutes. The task was available for completion for seven days, after which it expired. Upon completing the task, the author manually checks the responses for approval. Only participants in the United States with a HIT approval rate greater than 95% were eligible to participate.

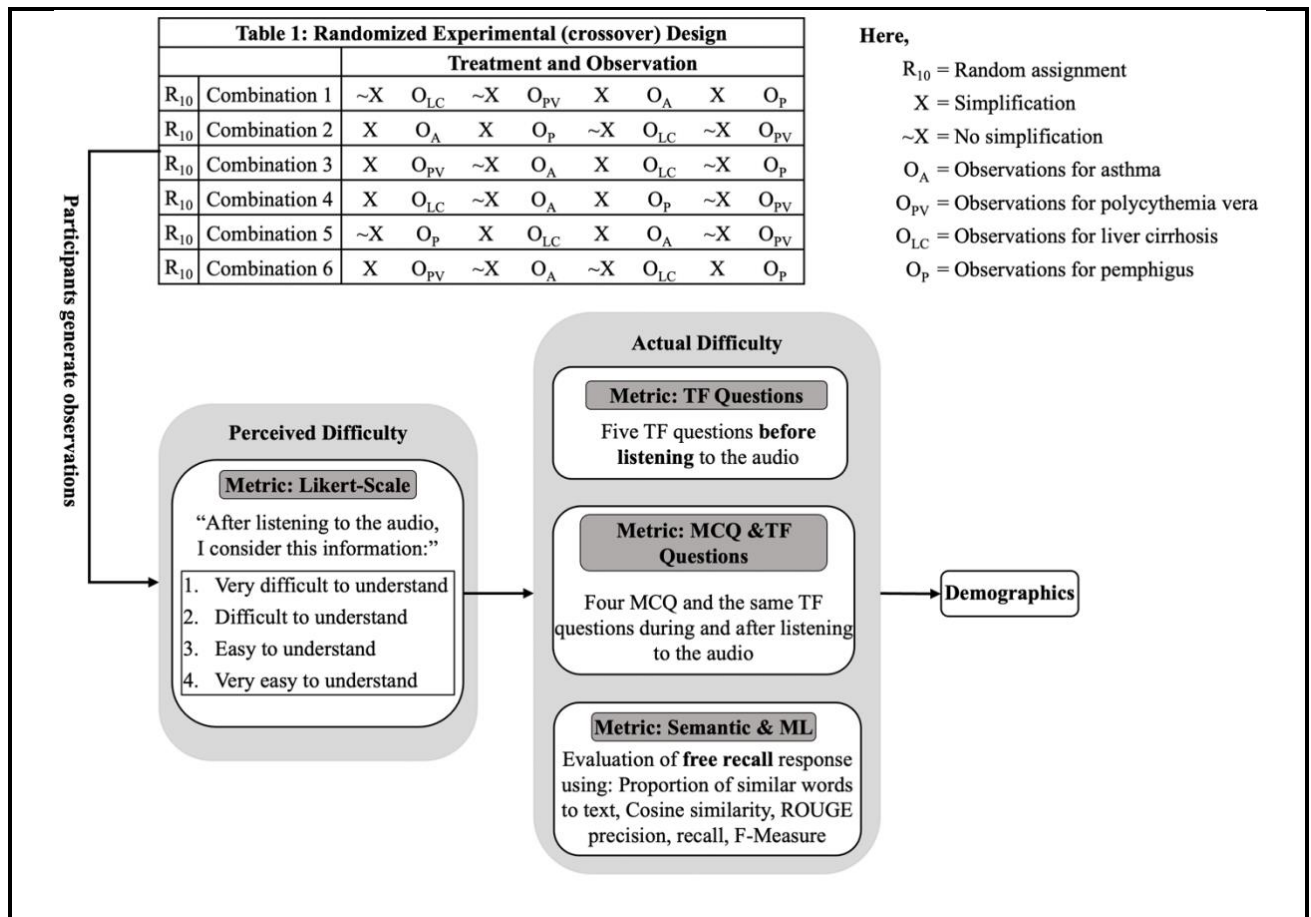

**Figure 2: Experimental Design and Research Methodology Flowchart**

## 2.4 Analytical Measures

In our study, we employed cosine similarity, a metric used to measure the cosine of the angle between two non-zero vectors, thereby quantifying the similarity between two pieces of text based on their vectorized word distributions (Singhal, 2001). Additionally, we applied automated evaluation metrics, specifically ROUGE (Recall-Oriented Understudy for Gisting Evaluation) measures, which assess the overlap between generated and reference summaries by comparing n-grams, word sequences, and word pairs (Lin, 2004). The study also evaluated the difficulty of the audio content using a 4-point Likert scale, ranging from "Very difficult to understand" to "Very easy to understand." Consistent with Fowler's guidelines (Fowler Jr, 2013), we adopted this ordinal scale to capture more detailed factual data, thereby enhancing the granularity of both perceived and actual difficulty assessments, as suggested by (Bhattacherjee, 2012).

We reused the multiple-choice questions (MCQ) and free recall measures. For multiple-choice questions, the answers were randomized. There were eleven questions. First, five true/false (TF) questions were presented before and after listening to the audio content. These questions allowed us to measure the learning of the basic concept presented in the audio. Second, six multiple-choice questions were presented during and after participants finished the audio. One attention check question and one general concept question were also presented. Finally, we presented a free recall question. We collected free recall information from participants after they listened to each audio. We calculated cosine similarities between free-recall responses and actual text to measure the semantic match between participants' answers and the text. We used cosine similarity based on Word2Vec word embeddings (Mikolov et al., 2013) vector representations of words. To compare the results of our study with the original study, we used an agreement column (**Agreement with original study**) in Table 4 based on the statistical significance.

We also implemented ROUGE measures (Lin, 2004) to assess the participants' information recall by evaluating the recall, precision, and F-measure of the longest phrases, unigrams, and bigrams compared to the reference text (text used to generate audio). Here, F-measures is the harmonic mean of precision and recall, and unigram bigram is the individual and pairs of words in a text, respectively.

# 3 Results

## 3.1 Participants and Descriptive Statistics

Sixty AMT participants completed the study and completed 240 audios. To ensure that the data analyzed was based on genuine effort and engagement by the participants, we excluded submissions where the participant spent <1 minute listening to the audio content and answering the accompanying multiple-choice questions in more than one section. We used the amount of time spent reading the text as an indicator of the participant's actual intent. The cutoff of one minute was chosen because it was the fastest the author could listen to the content (fast-forwarding audio) and answer questions while still providing thoughtful responses. We also excluded participants who provided the wrong completion code. From the four sections, we excluded the portion where participants did not pass the attention check question. This ensured a fair exclusion criterion that maintained a balance without being too lenient or stringent. After cleaning the data, 44 participants and 157 condition-based responses were retained. The demographic characteristics of the 44 participants are reported in Table 2.

Table 2: Participant Demographic Characteristics.

| Variable | Choice | Replication Count (%) | Original Count (%) |
|---|---|---|---|
| Sex | Female | 17 (39) | 39 (87) |
| | Male | 27 (61) | 6 (13) |
| Race (multiple options possible) | American Indian or Alaska Native | 3 (7) | 5 (11) |
| | Asian | 0 (0) | 1 (2) |
| | Black or African-American | 2 (5) | 1 (2) |
| | Native Hawaiian or Other Pacific Islander | 0 (0) | 1 (2) |
| | White | 39 (89) | 40 (89) |
| Ethnicity | Hispanic or Latino | 7 (16) | 19 (42) |
| | Not Hispanic or Latino | 37 (84) | 26 (58) |
| Education level | Less than a high school diploma | 1 (2) | 1 (2) |
| | High school diploma | 2 (5) | 11 (24) |
| | Associate degree | 1 (2) | 7 (16) |
| | Bachelor's degree | 6 (14) | 13 (29) |
| | Master's degree | 34 (77) | 11 (24) |
| | Doctoral degree (PhD, MD, ...) | 0 (0) | 2 (4) |
| Age | <=30 | 17 (39) | 10 (22) |
| | 31-40 | 20 (45) | 8 (18) |
| | 41-50 | 5 (11) | 9 (20) |
| | >50 | 2 (5) | 18 (40) |
| Language spoken at home | Never English | 0 (0) | 0 (0) |
| | Rarely English | 7 (16) | 1 (2) |
| | Half English | 1 (2) | 3 (7) |
| | Mostly English | 24 (55) | 14 (31) |
| | Only English | 12 (27) | 27 (60) |
| Consumption | Mostly verbal | 25 (57) | NA |
| | Mostly nonverbal | 9 (20) | NA |
| | Both | 10 (23) | NA |

## 3.2 Perceived Difficulty

Table 3 presents the one-way ANOVA results of the perceived difficulty scores between simplified and original audio. In this replication study, participants perceived the simplified content as significantly (F-value: 14.460, p: 0.001) more understandable (Mean: 2.98, SD: 0.694) than the original content (Mean: 2.52, SD: 0.833). Similarly, in the original study, the simplified content was also rated as significantly more comprehensible (Mean: 2.55, SD: 0.681) than the original content (Mean: 2.28, SD: 0.697), with a statistically significant difference (F-value: 6.130, p: 0.014). These findings underscore the text simplification's consistent impact on perceived difficulty in both studies.

Table 3: One-way ANOVA for Perceived Difficulty Scores

| Study | Metric | Condition | | F/ P- value |
|---|---|---|---|---|
| | | Original Mean (SD) | Simplified Mean (SD) | |
| Replication | 4-point Likert Scale | 2.52 (0.833) | 2.98 (0.694) | 14.460/ <0.001 |
| Original | 4-point Likert Scale | 2.28 (0.697) | 2.55 (0.681) | 6.130/ 0.014 |

*Comment: Both the original and replication studies found a significant perceived difference between the original and the simplified content.*

## 3.3 Actual Difficulty

Table 4 compares the actual difficulty levels between the original and replication studies across multiple metrics, including mean values, standard deviations (SD), and statistical significance. Following the original study, we conducted a one-way ANOVA for each item separately, with the results presented alongside a comparison to the original study.

Table 4: ANOVA Results of Actual Difficulty.

| Variable | Metric | Replication | | | Original | | | Agreement with the original study |
|---|---|---|---|---|---|---|---|---|
| | | Original Mean (SD) | Simplified Mean (SD) | F/P-Value | Original Mean (SD) | Simplified Mean (SD) | F/P-Value | |
| Questions before listening | TF1 | 37.12 (40.03) | 61.64 (22.41) | 15.342 /0.001 | 67 (47) | 63 (49) | 0.261/ 0.610 | Disagree |
| | TF2 | 72.73 (28.51) | 73.16 (28.26) | 0.05 /0.823 | 71 (46) | 71 (46) | 0.001/ 0.982 | Agree |
| | TF3 | 43.86 (12.94) | 55.64 (8.87) | 2.335 /0.129 | 51 (50) | 53 (50) | 0.061/ 0.805 | Agree |
| | TF4 | 79.39 (18.55) | 85.28 (16.25) | 0.706 /0.402 | 76 (43) | 72 (45) | 0.257/ 0.613 | Agree |
| | TF5 | 36.36 (11.3) | 41 (11.6) | 0.802 /0.372 | 33 (47) | 59 (50) | 11.450/ 0.001 | Disagree |
| Questions during & after listening | General Question | 45.98 (25.15) | 37.92 (17.89) | 1.154 /0.284 | 71 (79) | 70 (76) | 0.024/ 0.877 | Agree |
| | MCQ1 | 34.85 (19.63) | 39 (28.34) | 0.001 /0.977 | 37 (48) | 57 (50) | 6.320/ 0.013 | Disagree |
| | MCQ2 | 38.71 (24.79) | 43.39 (19.61) | 0.001 /0.971 | 68 (47) | 75 (44) | 0.835/ 0.362 | Agree |
| | MCQ3 | 40.3 (24.15) | 48.25 (19.33) | 0.093 /0.761 | 33 (47) | 59 (50) | 11.450/ 0.001 | Disagree |
| | MCQ4 | 32.42 (23.72) | 71.6 (11.07) | 6.068 /0.015 | 33 (47) | 29 (46) | 0.282/ 0.596 | Disagree |
| | TF1 | 46.36 (30.31) | 65.7 (20.27) | 8.193 /0.005 | 66 (48) | 62 (49) | 0.263/ 0.609 | Disagree |
| | TF2 | 70 (24.08) | 74.22 (24.35) | 0.868 /0.353 | 93 (27) | 88 (32) | 0.793/ 0.375 | Agree |
| | TF3 | 37.2 (14.82) | 57.31 (11.31) | 6.602 /0.011 | 66 (48) | 72 (45) | 0.770/ 0.382 | Disagree |
| | TF4 | 83.18 (15.14) | 80.81 (20.44) | 0.038 /0.847 | 93 (27) | 92 (27) | 0.005/ 0.945 | Agree |
| | TF5 | 44.55 (8.98) | 52.45 (13.4) | 1.53 /0.218 | 52 (50) | 78 (42) | 11.921/ 0.001 | Disagree |
| Semantic evaluation | Proportion of similar words to text | 0.054 (0.046) | 0.071 (0.054) | 4.333/0.039 | 76 (16) | 82 (11) | 6.729/ 0.010 | Agree |
| | Cosine similarity | 0.409 (0.189) | 0.460 (0.181) | 2.88/0.092 | 0.111 (0.013) | 0.116 (0.009) | 9.207/ 0.003 | Disagree |
| ML evaluation | ROUGE Recall (Longest Phrase) | 0.007 (0.022) | 0.008 (0.013) | 0.279/0.598 | 0.093 (0.064) | 0.108 (0.070) | 1.894/ 0.171 | Agree |
| | ROUGE Precision (Longest Phrase) | 0.048 (0.115) | 0.065 (0.102) | 1.012/0.316 | 0.487 (0.212) | 0.436 (0.125) | 3.380/ 0.068 | Agree |
| | ROUGE F-Measure (Longest Phrase) | 0.012 (0.037) | 0.014 (0.021) | 0.331/0.566 | 0.151 (0.095) | 0.164 (0.090) | 0.678/ 0.412 | Agree |
| | ROUGE Recall (Unigram) | 0.059 (0.052) | 0.078 (0.061) | 4.115/0.044 | 0.085 (0.05) | 0.119 (0.080) | 9.825/ 0.002 | Agree |
| | ROUGE Precision (Unigram) | 0.479 (0.209) | 0.520 (0.189) | 1.56/0.213 | 0.689 (0.152) | 0.654 (0.133) | 2.424/ 0.122 | Agree |
| | ROUGE F-Measure (Unigram) | 0.100 (0.078) | 0.128 (0.091) | 4.307/0.04 | 0.146 (0.079) | 0.191 (0.111) | 8.263/ 0.005 | Agree |
| | ROUGE Recall (Bigram) | 0.017 (0.026) | 0.022 (0.022) | 1.348/0.247 | 0.037 (0.038) | 0.044 (0.039) | 1.221/ 0.271 | Agree |
| | ROUGE Precision (Bigram) | 0.176 (0.148) | 0.202 (0.138) | 1.383/0.241 | 0.316 (0.263) | 0.240 (0.145) | 4.841/ 0.029 | Disagree |
| | ROUGE F-Measure (Bigram) | 0.031 (0.043) | 0.038 (0.035) | 1.42/0.235 | 0.064 (0.064) | 0.070 (0.056) | 0.378/ 0.540 | Agree |

### 3.4 Results Before Listening to Audio

In response to the questions presented before listening to the audio content, participants' performance varied across the TF questions. For TF1, participants exposed to the simplified audio content had a higher mean accuracy (Mean: 61.64, SD: 22.41) compared to those exposed to the original audio (Mean: 37.12, SD: 40.03), with a statistically significant difference (F-value: 15.342, $p < 0.001$). For TF2, there was no significant difference between the simplified (Mean: 73.16, SD: 28.26) and original audio (Mean: 72.73, SD: 28.51; F-value: 0.05, $p = 0.823$). Similarly, TF3 and TF4 showed no significant differences between the simplified and original audio (TF3: F-value: 2.335, $p = 0.129$; TF4: F-value: 0.706, $p = 0.402$).

The original study found no significant difference for TF1 (F-value: 0.261, $p = 0.610$). However, it reported a significant difference for TF5 (F-value: 11.450, $p = 0.001$), which was not observed in the replication study (F-value: 0.802, $p = 0.372$).

### 3.5 Results During and After Listening to Audio

For the multiple-choice questions (MCQs), participants' performance varied across the questions. MCQ1, MCQ2, and MCQ3 showed no significant differences between the simplified and original audio. However, MCQ4 revealed a statistically significant improvement in accuracy for the simplified audio (Mean: 71.60, SD: 11.07) compared to the original audio (Mean: 32.42, SD: 23.72; F-value: 6.068, $p = 0.015$).

For the TF questions presented after listening to the audio, participants demonstrated significant improvements in accuracy for TF1 and TF3 when exposed to the simplified audio. For TF1, the simplified audio resulted in a higher mean accuracy (Mean: 65.70, SD: 20.27) than the original audio (Mean: 46.36, SD: 30.31; F-value: 8.193, p = 0.005). Similarly, for TF3, the simplified audio showed higher mean accuracy (Mean: 57.31, SD: 11.31) than the original audio (Mean: 37.20, SD: 14.82; F-value: 6.602, p = 0.011). However, no significant differences were observed for TF2 (F-value: 0.868, p = 0.353), TF4 (F-value: 0.038, p = 0.847), or TF5 (F-value: 1.53, p = 0.218).

In comparison to the original study, the replication study found significant improvements for MCQ4, TF1, and TF3, whereas the original study reported significant improvements for MCQ1 (F-value: 6.320, p = 0.013) and MCQ3 (F-value: 11.450, p = 0.001). The original study also found significant differences for TF5 (F-value: 11.921, p = 0.001), which was not observed in the replication study.

### 3.6 Results of Semantic and ML Evaluation

For the semantic evaluation of participants' free recall responses, the proportion of words similar to the text was significantly higher for the simplified audio (Mean: 0.071, SD: 0.054) compared to the original audio (Mean: 0.054, SD: 0.046; F-value: 4.333, p = 0.039). However, the cosine similarity between participants' responses and the text, while higher for the simplified audio (Mean: 0.460, SD: 0.181) compared to the original (Mean: 0.409, SD: 0.189), was not statistically significant (F-value: 2.88, p = 0.092). The original study reported significant differences in cosine similarity (F-value: 9.207, p = 0.003) and ROUGE Precision (Bigram; F-value: 4.841, p = 0.029).

For the ROUGE measures, ROUGE Recall (Unigram) showed a significant improvement for the simplified audio (F-value: 4.115, p = 0.044), as did the F-measure (Unigram; F-value: 4.307, p = 0.040). Similarly, the original study reported significant improvement in ROUGE precision and recall (unigram) measures.

The replication study found no significant improvements in cosine similarity compared to the original study, but the original study did (F-value: 9.207, p = 0.003).

## 4 Secondary Analysis

Following the original study, we conducted two supplementary analyses on self-reported data: educational attainment and language proficiency. Education level was categorized on a scale from 1 to 6: *1 = Less than a high school diploma, 2 = High school diploma, 3 = Associate degree, 4 = Bachelor's degree, 5 = Master's degree, and 6 = Doctoral degree (PhD, MD, etc.)*. Similarly, language proficiency was measured on a scale from 1 to 5 based on the primary language spoken at home: *1 = Never English, 2 = Rarely English, 3 = Half English, 4 = Mostly English, and 5 = Only English*. Our analysis showed that education level and language proficiency were related to participants' accuracy in answering questions, though in different directions (see Figure 3). Education level was significantly negatively correlated with accuracy ($\rho$ = -0.173, T-value = -2.177, p = 0.030). Compared to the original study, they found a significant positive correlation with answers accuracy in both cases. In both studies, simplified content improved the accuracy of answering questions regardless of education, language, and text/ audio length.

In our study, the accuracy of answering questions increased with the level of education initially, but later, it showed a consistent decrease. However, the original study did not observe this trend. The percentage accuracy of answering questions increased in both studies with the higher level of English language proficiency. We also analyzed how other demographic characteristics might affect performance. We analyzed additional demographic information such as age, gender, race, ethnicity, and media consumption habits. We calculated the average scores and standard deviations for each demographic group. However, we found no significant differences between these groups (Appendix A, Table 1).

## 5 Discussion

This methodological replication study investigated the impact of text simplification on comprehension of healthcare information presented in audio format, extending the original study's findings, which focused on written text. Our results largely corroborate the effectiveness of text simplification across modalities while also revealing some differences that demand further exploration.

The overall impact of text simplification on comprehension aligns with the original study's findings, demonstrating improved performance across various metrics for simplified content. This consistency suggests that the benefits of text simplification are not confined to visual presentation but extend to auditory information. Such robustness across modalities underscores the potential of text simplification to enhance healthcare communication.

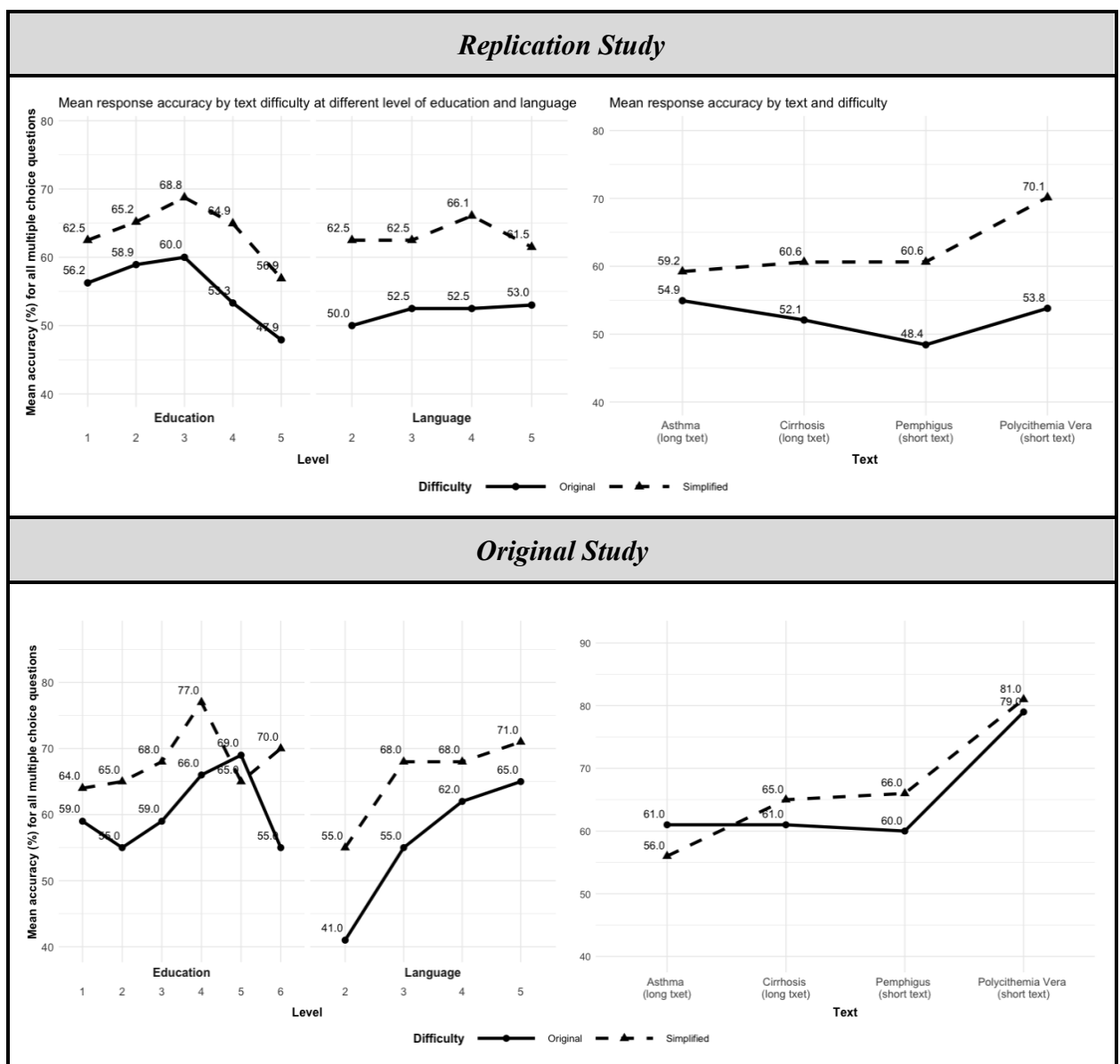

**Figure 3: Mean accuracy answering questions by education, language level, and text for the original and replication study.**

At first glance, our study appears to reveal notable differences in the effectiveness of simplification across question types compared to the original study. However, a closer inspection reveals that both studies point in the same direction, particularly for True/False (TF) questions. Both studies observed overall performance improvement after participants were exposed to the content, with even higher improvements for simplified content. Before content exposure, while there were apparent disagreements in TF performance, both studies found one significant performance improvement each (replication: TF1; original: TF5). Importantly, when presented with the same TF questions after consuming the content, both studies found the same TF questions to be significant respectively. This consistency suggests that simplification benefits comprehension across modalities, even if the specific questions showing improvement may vary. The results for Multiple Choice Questions (MCQs) showed a similar pattern of underlying agreement despite surface-level differences. Our study found significant improvement for MCQ4 with simplified audio, while the original study showed significant improvements for MCQ1 and MCQ3. These apparent discrepancies may be attributed to the change in modality from text to audio, rather than fundamental differences in the effectiveness of simplification. In text format, participants can read and comprehend at their own pace, potentially revisiting complex information. However, factors such as emphasis, pacing, and strategic pauses are essential for audio content to increase comprehension and retention (Ahmed et al., 2024). The significant differences in results across MCQs may reflect these modality-specific information processing and recall challenges.

Interestingly, both studies observed decreased performance for the general question with simplified content, although this decrease was not statistically significant. This consistent finding across modalities suggests that simplification may not always benefit broader, more open-ended questions about the content. It's possible that simplified language while aiding in the comprehension of specific details, may inadvertently reduce the richness of context needed to answer more general questions.

Our study extended the analysis to examine the roles of education level and language proficiency in comprehension outcomes, finding that higher education levels and English proficiency positively correlated with improved accuracy in answering questions. Importantly, simplification benefited participants across all education levels, highlighting its potential as a universal strategy for enhancing healthcare communication. These findings emphasize the need for tailored approaches to health literacy interventions that consider the diverse backgrounds of the target audience.

The differences observed between audio and written presentations suggest that healthcare communicators should carefully consider the modality of information delivery when applying simplification techniques. While simplification generally improves comprehension across modalities, the specific benefits may vary depending on the content type and assessment method. Future research should explore why certain question types or content areas benefit more from simplification in audio versus written formats and investigate strategies to mitigate the potential challenges associated with audio presentation, such as increased cognitive load and limitations on information review.

## 6  Limitations

While our study aimed to provide insights into the effectiveness of simplified medical audio, several limitations are acknowledged. This study replicates the original methodology but focuses on audio content, incorporating Microsoft Azure for text-to-speech conversion, which introduces a new variable. A key limitation of using Azure is the challenge of standardizing audio parameters to meet all listeners' diverse preferences and comprehension needs. Specifically, using an American English male voice may introduce a bias towards a particular linguistic and gendered perspective. Although we adjusted parameters like speech rate, pauses, and pronunciation to improve audio quality, individual preferences for voice styles vary significantly. Nevertheless, the use of Azure did not substantially affect our main findings, as the primary goal of this replication was to compare comprehension levels between original and simplified audio.

Another limitation is the absence of benchmarks using advanced AI models like ChatGPT (Brown, 2020). At the time of our research, these models had not yet gained widespread adoption or integration into mainstream text simplification tools. Consequently, our study did not include a comparison with these models. Future research could address this gap by evaluating the performance of text and audio simplification using more recent AI advancements, potentially offering improved capabilities and additional insights into the effectiveness of various simplification methods (Leroy et al., 2024).

Lastly, we did not include a manual evaluation of participants' free recall responses, as in the original study. Instead, we relied on automated, scalable techniques better suited for handling large data volumes. This approach is consistent with recent studies prioritizing automation to minimize bias and inconsistency (Manzoor et al., 2021). However, this choice may have limited the qualitative depth of our findings, potentially missing subtle nuances in participants' responses.

## 7  Conclusion

Comprehending medical information is difficult for laypersons due to the complexity of medical terminology and jargon and the technical nature of medical concepts and procedures. This replication study reaffirms the impact of text simplification on enhancing healthcare information comprehension for both text and audio content. The findings consistently demonstrate that simplified healthcare information is perceived as more understandable and improves comprehension, aligning with the original study's observations. This underscores the practical utility of text simplification tools in promoting health literacy and ensuring that healthcare information reaches a broader audience.

Moreover, the analysis of participants' education levels and language proficiency highlights the importance of tailoring health communication strategies to the specific needs and backgrounds of the target audience. Additionally, participants with stronger English language proficiency consistently outperformed their counterparts, emphasizing the role of language skills in healthcare information access. This research reestablished the value of text simplification as a versatile tool for improving healthcare information dissemination in audio format, with the caveat that a nuanced approach, considering factors such as education and language proficiency, can further enhance its effectiveness.

## Acknowledgments

The research reported in this paper was supported by the National Library of Medicine of the National Institutes of Health under Award Number R01LM011975. The content is solely the authors' responsibility and does not necessarily represent the official views of the National Institutes of Health.

14                                                                                    {Article Title}Wulf, Linda, et al. "Hands free-care free: elderly people taking advantage of speech-only interaction." Proceedings of the 8th Nordic Conference on Human-Computer Interaction: Fun, Fast, Foundational. 2014.

Bai, Zechen, et al. "Hallucination of multimodal large language models: A survey." arXiv preprint arXiv:2404.18930 (2024).

Dong, Yi, et al. "Safeguarding Large Language Models: A Survey." arXiv preprint arXiv:2406.02622 (2024).

Williams, Mark V., et al. "Relationship of functional health literacy to patients' knowledge of their chronic disease: a study of patients with hypertension and diabetes." Archives of internal medicine 158.2 (1998): 166-172.

Lin, Chin-Yew. "Rouge: A package for automatic evaluation of summaries." In Text summarization branches out, pp. 74-81. 2004.

Brown, Tom B. "Language models are few-shot learners." arXiv preprint arXiv:2005.14165 (2020).

Leroy, Gondy, David Kauchak, Philip Harber, Ankit Pal, and Akash Shukla. "Text and Audio Simplification: Human vs. ChatGPT." AMIA Summits on Translational Science Proceedings 2024 (2024): 295.
Volume YY                                                                                    Paper XX

# Appendix A: Additional Secondary Analysis

**Table 1: Demographic Specific Actual and Perceived Difficulty Analysis.**

| Variable | Choice | Actual Difficulty | | Perceived Difficulty | |
| --- | --- | --- | --- | --- | --- |
| | | Original Mean % (SD) | Simplified Mean % (SD) | Original | Simplified |
| Age | <=30 | 54.23 (14.47) | 59.76 (13.91) | 2.58 (0.81) | 3.05 (0.67) |
| | 31-40 | 50.48 (10.89) | 66.21 (14.01) | 2.58 (0.86) | 2.94 (0.72) |
| | 41-50 | 54.86 (22.70) | 63.89 (16.76) | 2.22 (0.83) | 2.78 (0.83) |
| | >50 | 52.08 (9.55) | 53.13 (27.72) | 2.33 (1.15) | 3.25 (0.50) |
| Gender | Female | 54.02 (16.42) | 59.58 (15.98) | 2.61 (0.88) | 3.10 (0.66) |
| | Male | 51.18 (12.91) | 63.77 (13.59) | 2.46 (0.80) | 2.91 (0.71) |
| Race | American Indian or Alaska Native | 60.16 (6.63) | 61.72 (18.43) | 2.75 (0.71) | 3.25 (0.46) |
| | Asian | 0.00 (0.00) | 0.00 (0.00) | 0.00 (0.00) | 0.00 (0.00) |
| | Black or African-American | 68.75 (8.84) | 76.56 (13.86) | 2.00 (0.82) | 3.25 (0.50) |
| | Native Hawaiian or Other Pacific Islander | 0.00 (0.00) | 0.00 (0.00) | 0.00 (0.00) | 0.00 (0.00) |
| | White | 50.66 (14.35) | 61.57 (14.66) | 2.53 (0.85) | 2.95 (0.72) |
| Ethnicity | Hispanic or Latino | 50.63 (9.97) | 70.59 (13.76) | 2.70 (0.67) | 3.00 (0.50) |
| | Not Hispanic or Latino | 52.73 (15.20) | 60.17 (14.05) | 2.49 (0.86) | 2.97 (0.74) |
| Consumption | Mostly verbal | 51.14 (11.21) | 57.14 (13.35) | 2.73 (0.72) | 3.12 (0.67) |
| | Mostly nonverbal | 50.39 (16.37) | 66.15 (12.62) | 2.31 (0.87) | 2.83 (0.70) |
| | Both | 57.03 (18.10) | 69.10 (15.82) | 2.31 (0.95) | 2.83 (0.71) |

# Appendix B: Text and Questions

The four medical texts to generate the audio and the instrument set to assess the perceived and actual difficulty were borrowed from the original study by Leroy et al. (2022). All the material presented in this appendix was taken from the appendix of the original study.

**Text: Asthma**

**True/False Questions Before Reading the Text**

The severity of asthma is determined by multiple tests. [[true]]

Taking asthma medication for prolonged periods can make asthma worse. [[false]]

Having hay fever is a main cause of asthma in many people. [[false]]

Research has shown that asthma is linked to obesity. [[true]]

Corticosteroids are the only effective inhaled medication for asthma. [[false]]

**Original Text**

Asthma (from the Greek ἆσθμα, ásthma, "panting") is the common chronic inflammatory disease of the airways characterized by variable and recurring symptoms, reversible airflow obstruction, and bronchospasm. Symptoms include wheezing, coughing, chest tightness, and shortness of breath. Asthma is clinically classified according to the frequency of symptoms, forced expiratory volume in 1 second (FEV1), and peak expiratory flow rate. Asthma may also be classified as atopic (extrinsic) or non-atopic (intrinsic).

It is thought to be caused by a combination of genetic and environmental factors. Treatment of acute symptoms is usually with an inhaled short-acting beta-2 agonist (such as salbutamol). Symptoms can be prevented by avoiding triggers, such as allergens and irritants, and by inhaling corticosteroids. Leukotriene antagonists are less effective than corticosteroids and thus less preferred.

Its diagnosis is usually made based on the pattern of symptoms and/or response to therapy over time. The prevalence of asthma has increased significantly since the 1970s. As of 2010, 300 million people were affected worldwide. In 2009 asthma caused 250,000 deaths globally. Despite this, with proper control of asthma with step down therapy, prognosis is generally good.

Causes

Asthma is caused by environmental and genetic factors. These factors influence how severe asthma is and how well it responds to medication. The interaction is complex and not fully understood.

Studying the prevalence of asthma and related diseases such as eczema and hay fever have yielded important clues about some key risk factors. The strongest risk factor for developing asthma is a history of atopic disease; this increases one's risk of hay fever by up to 5× and the risk of asthma by 3–4×. In children between the ages of 3–14, a positive skin test for allergies and an increase in immunoglobulin E increases the chance of having asthma. In adults, the more allergens one reacts positively to in a skin test, the higher the odds of having asthma.

Because much allergic asthma is associated with sensitivity to indoor allergens and because Western styles of housing favor greater exposure to indoor allergens, much attention has focused on increased exposure to these allergens in infancy and early childhood as a primary cause of the rise in asthma. Primary prevention studies aimed at the aggressive reduction of airborne allergens in a home with infants have shown mixed findings. Strict reduction of dust mite allergens, for example, reduces the risk of allergic sensitization to dust mites, and modestly reduces the risk of developing asthma up until the age of 8 years old. However, studies also showed that the effects of exposure to cat and dog allergens worked in the converse fashion; exposure during the first year of life was found to reduce the risk of allergic sensitization and of developing asthma later in life.

The inconsistency of this data has inspired research into other facets of Western society and their impact upon the prevalence of asthma. One subject that appears to show a strong correlation is the development of asthma and obesity. In the United Kingdom and United States, the rise in asthma prevalence has echoed an almost epidemic rise in the prevalence of obesity. In Taiwan, symptoms of allergies and airway hyper-reactivity increased in correlation with each 20% increase in body-mass index. Several factors associated with obesity may play a role in the pathogenesis of asthma, including decreased respiratory function due to a buildup of adipose tissue (fat) and the fact that adipose tissue leads to a pro-inflammatory state, which has been associated with non-eosinophilic asthma.

Asthma has been associated with Churg–Strauss syndrome, and individuals with immunologically mediated urticaria may also experience systemic symptoms with generalized urticaria, rhino-conjunctivitis, orolaryngeal and gastrointestinal symptoms, asthma, and, at worst, anaphylaxis. Additionally, adult-onset asthma has been associated with periocular xanthogranulomas.

**Simplified Text**

Asthma (from the Greek word for "panting," which means rapid shallow breathing) is the common chronic inflammatory condition of the airways characterized by recurring symptoms, blocking of the air flow, and spasming of the airways (bronchospasms). Symptoms include wheezing, coughing, tightening of the chest, and shortness of breath. The severity of asthma is determined based on how often symptoms happen and the results of two tests: 1) the amount of air blown in 1 second (FEV1), and the strongest amount of air that can be blown. Asthma may also be classified as atopic (outside) or non-atopic (inside).

Asthma is thought to be caused by a combination of genetic and environmental factors. Treatment of is usually an inhaled medicine (such as salbutamol). Symptoms can be prevented by avoiding triggers, such as allergens and irritants, and by inhaling medicine (for example, corticosteroids). For medicine, corticosteroids are more effective than leukotriene antagonists.

Diagnosis of asthma is usually made based on the pattern of symptoms and/or response to therapy over time. The number of people with asthma has increased significantly since the 1970s. As of 2010, 300 million people had it worldwide. In 2009, 250,000 worldwide died of asthma. However, with proper treatment, outcomes are generally good.

Causes

Asthma is caused by both the environment and genes. The cause determines how serious asthma is and how well it responds to medicine. How the genes and environment interact is complex and not fully understood.

Studying the prevalence of asthma and related conditions such as eczema and hay fever have given important clues about some key risk factors. The strongest risk factor for developing asthma is a history of reacting to things in the environment. The risk of having hay fever increases by up to 5 times and the risk of asthma by 3–4 times. In children between the ages of 3 to 14, a positive skin test for allergies and an increase in immunoglobulin E increases the chance of getting asthma. In adults, the more allergens one reacts to in a skin test, the higher the odds of having asthma.

Allergic asthma is linked with sensitivity to indoor allergens, and Western styles of housing have more indoor allergens. Research has focused on increased exposure to these allergens when people are young as a main cause of the increases in asthma. Studies aimed at preventing asthma looked at whether greatly reducing airborne allergens in a home with babies have shown mixed findings. Reducing dust mite allergens, for example, cut down the risk of allergy to dust mites, and modestly cut the risk of developing asthma up until the age of 8 years old. However, studies also showed that exposure to cats and dogs was beneficial. Exposure during the first year of life to cats and dogs was found to cut down the risk of allergic sensitization and of developing asthma later in life.

The inconsistency of this data has resulted in research into other parts of Western society and their impact upon the prevalence of asthma. One subject that seems to correlate strongly is the development of asthma and obesity. In the United Kingdom and United States, the increase in asthma has mirrored the increase in obesity. In Taiwan, symptoms of allergies and asthma increased with each 20% increase in body-mass index. Several factors linked with obesity may play a part in the beginning of asthma, including reduced respiratory function due to a buildup of fat and the fact that fatty tissue results in a pro-inflammatory state, which has been linked with asthma.

Asthma has been linked with Churg–Strauss syndrome, and people who get hives may also experience other symptoms: rhino-conjunctivitis, orolaryngeal and gastrointestinal symptoms, asthma, and, at worst, allergic shock. Adult asthma has been linked with periocular xanthogranulomas.

**Multiple-choice Questions Presented with the Text**

(perceived difficulty) After reading this text, I consider this information:

Very difficult to understand

Difficult to understand

Easy to understand

Very easy to understand

(Overview question) This text provides information on risk factors for asthma:

No, it mainly discusses the different causes of asthma.

No, it mainly discusses different types of asthma

Yes, it discusses allergens as risk factors. [[answer]]

Yes, it discusses childhood presence of pets as risk factors.

(General question) Asthma is caused

by environmental and genetic factors. [[answer]]

mostly by environmental factors.

mostly by genetic factors.

by triggers in the environment.

**Multiple-choice Questions Presented after Reading the Text**

Asthma symptoms include all of the following *except*:

coughing.

slow breathing. [[answer]]

wheezing.

tightening in the chest.

How are asthma symptoms most effectively avoided?

by avoiding triggers and using inhaled medication. [[answer]]

by using inhaled medication.

by using inhaled medication once symptoms occur.

by avoiding triggers.

Which of the following is a common measure for diagnosing asthma?

the amount of air you can breathe in in one second.

the amount of air you can breathe out in one second. [[answer]]

the amount of air you can breathe in in one breath.

the amount of air you can breathe out in one breath.

Which of the following is NOT related to developing asthma?

obesity.

hay fever. [[answer]]

exposure to dogs and cats at a young age.

dust mite allergens at a young age.

**Repeat True/False Questions Presented after Reading the Text**

The severity of asthma is determined by multiple tests. [[true]]

Taking asthma medication for prolonged periods can make asthma worse. [[false]]

Having hay fever is a main cause of asthma in many people. [[false]]

Research has shown that asthma is linked to obesity. [[true]]

Corticosteroids are the only effective inhaled medication for asthma. [[false]]

**Text: Liver cirrhosis**

**True/False Questions before Reading the Text**

Liver cirrhosis can be cured if caught very early on. [[false]]

Liver cirrhosis can be caused by alcoholism or hepatitis. [[true]]

The liver is an essential organ because it generates red blood cells. [[false]]

People with liver cirrhosis can have both scars and fibrous tissue in the liver. [true]

A visual inspection of the liver using laparoscopy is commonly used for diagnosis.  [[false]]

**Original Text**

Cirrhosis is a consequence of chronic liver disease characterized by replacement of liver tissue by fibrosis, scar tissue and regenerative nodules (lumps that occur as a result of a process in which damaged tissue is regenerated), leading to loss of liver function. Cirrhosis is most commonly caused by alcoholism, hepatitis B and C, and fatty liver disease, but has many other possible causes. Some cases are idiopathic (i.e., of unknown cause).

Ascites (fluid retention in the abdominal cavity) is the most common complication of cirrhosis, and is associated with a poor quality of life, increased risk of infection, and a poor long-term outcome. Other potentially life-threatening complications are hepatic encephalopathy (confusion and coma) and bleeding from esophageal varices. Cirrhosis is generally irreversible, and treatment usually focuses on preventing progression and complications. In advanced stages of cirrhosis, the only option is a liver transplant.

The word "cirrhosis" derives from Greek κιρρός [kirrhós] meaning yellowish, tawny (the orange-yellow colour of the diseased liver) + Eng. med. suff. -osis. While the clinical entity was known before, it was René Laennec who gave it the name "cirrhosis" in his 1819 work in which he also describes the stethoscope.

Pathophysiology

The liver plays a vital role in synthesis of proteins (e.g., albumin, clotting factors and complement), detoxification and storage (e.g., vitamin A). In addition, it participates in the metabolism of lipids and carbohydrates.

Cirrhosis is often preceded by hepatitis and fatty liver (steatosis), independent of the cause. If the cause is removed at this stage, the changes are still fully reversible.

The pathological hallmark of cirrhosis is the development of scar tissue that replaces normal parenchyma, blocking the portal flow of blood through the organ and disturbing normal function. Recent research shows the pivotal role of the stellate cell, a cell type that normally stores vitamin A, in the development of cirrhosis. Damage to the hepatic parenchyma leads to activation of the stellate cell, which becomes contractile (called myofibroblast) and obstructs blood flow in the circulation. In addition, it secretes TGF-β1, which leads to a fibrotic response and proliferation of connective tissue. Furthermore, it secretes TIMP 1 and 2, naturally occurring inhibitors of matrix metalloproteinases, which prevents them from breaking down fibrotic material in the extracellular matrix.

The fibrous tissue bands (septa) separate hepatocyte nodules, which eventually replace the entire liver architecture, leading to decreased blood flow throughout. The spleen becomes congested, which leads to hypersplenism and increased sequestration of platelets. Portal hypertension is responsible for most severe complications of cirrhosis.

Diagnosis

The gold standard for diagnosis of cirrhosis is a liver biopsy, through a percutaneous, transjugular, laparoscopic, or fine-needle approach. A biopsy is not necessary if the clinical, laboratory, and radiologic data suggests cirrhosis. Furthermore, there is a small but significant risk to liver biopsy, and cirrhosis itself predisposes for complications due to liver biopsy. Ascites, low platelet count, and spider nevi are useful physical findings.

**Simplified text:**

Cirrhosis is a result of chronic liver disease that leads to the liver not being able to work correctly. Cirrhosis replaces healthy liver tissue with fibrosis, scar tissue, and lumps from regrown damaged tissue (regenerative nodules). Cirrhosis is usually caused by alcohol abuse, hepatitis B and C, and fatty liver disease, but has many other causes. Some cases do not have a known cause (called, idiopathic cases).

Fluid build up in the abdomen (ascites) is the most common complication of cirrhosis, and is linked with a poor quality of life, increased risk of infection, and a poor long-term outcome. Other potentially life-threatening complications are confusion and coma (hepatic encephalopathy) and bleeding from varicose veins in the esophagus. Cirrhosis cannot be reversed, and treatment usually focuses on preventing the cirrhosis from getting worse and developing complications. In advanced stages of cirrhosis, the only choice is a liver transplant.

The word "cirrhosis" comes from the Greek word kirrhós, meaning yellowish or tawny (the orange-yellow color of the diseased liver). While the disease was known before, it was René Laennec who gave it the name "cirrhosis" in 1819 in a work where he also describes the stethoscope.

Pathophysiology

The liver is critical in making proteins (e.g., albumin, clotting factors and complement), removing toxins from the blood, and storing nutrients (e.g., vitamin A). It also takes part in the processing of fats (lipids) and sugars (carbohydrates).

No matter what the cause, hepatitis and fatty liver (steatosis) are usually present before the liver develops cirrhosis. If the cause of the hepatitis or fatty liver is removed at this stage, the changes to the liver can still be reversed.

The best indicator of cirrhosis is the development of scar tissue that replaces healthy liver tissue, blocking the flow of blood through the liver and keeping it from working normally. Recent research shows the important role of the stellate cell, a cell type that normally stores vitamin A, in the development of cirrhosis. Damage to the healthy liver tissue triggers the stellate cell, which becomes contracted (called myofibroblast) and blocks blood flow in the circulation. In addition, it secretes TGF-β1, which leads to a an over-production of connective tissue (healthy tissue is replaced by tissue that can't do the same functions). What is more, it secretes TIMP 1 and 2, naturally occurring inhibitors of matrix metalloproteinases, which prevents them from breaking down fibrotic material in the structure around the cells.

The fibrous tissue bands (septa) separate liver cell nodules, which eventually replace the entire liver structure, leading to decreased blood flow throughout. The spleen becomes congested, which leads to an enlarged spleen and increased separation of platelets. Portal hypertension (high blood pressure in the vein to the liver) is responsible for the worst complications of cirrhosis.

Diagnosis

Cirrhosis is best diagnosed with a liver biopsy. A biopsy is not necessary if the clinical, laboratory, and radiologic data indicate cirrhosis. Furthermore, there is a small but significant risk to liver biopsy, and cirrhosis itself predisposes for complications due to liver biopsy. Ascites, low platelet count, and moles with spider-like veins are other useful physical findings.

**Multiple-choice Questions Presented with the Text**

(Perceived Difficulty) After reading this text, I consider this information

Very difficult to understand

Difficult to understand

Easy to understand

Very easy to understand

(Overview Question) This text explains

Different treatment options for liver cirrhosis versus hepatitis.

Different treatment options developed for liver cirrhosis over the last century.

Effects of liver cirrhosis on the liver [[answer]]

Effects of liver cirrhosis on blood cell counts.

(General question) Which of the following is a complication of liver cirrhosis?

Retaining too many fluids in the body, which increases the chance of infections [[answer]]

Retaining too much fat, resulting in fatty liver disease

Not being able to retain any or enough Vitamin A

Weakness of the esophageal veins leading to a contamination of the blood

**Multiple-choice Questions Presented after Reading the Text**

How are liver cirrhosis and hepatitis related?

Hepatitis is a disease that occurs often before cirrhosis and directly causes it. [[answer]]

Hepatitis is a disease that occurs often as a result of cirrhosis and is caused by it.

Hepatitis is a blood disease that indirectly causes cirrhosis by preventing the absorption of vitamin A.

Hepatitis is a cell disease that indirectly causes cirrhosis by changing the shape of stellate cells.

Why is blood flow in the liver an important problem with liver cirrhosis?

Scar tissue develops, which requires new veins to be formed.

There is a multiplication of connective veins which requires more blood flow.

Both scar formation and changes in liver cells lead to reduced blood flow. [[answer]]

Scar tissue develops which reduces the number of working veins and leads to reduced blood flow.

What is a complication of liver cirrhosis?

increased blood pressure in the veins leading to the liver [[answer]]

toxic waste in the blood

reduced vision and yellowing of the eyes

a contraction of the veins leading from the liver

Why is the formation of scar tissue an important complication of liver cirrhosis?

Scar tissue replaces parts of the liver and activates stellate cells both of which lead to the blocking of blood flow through the liver [[answer]]

Scar tissue leads to obstructions in veins which leads to blocked blood flow

Fibrous veins are replaced by scar tissue which leads to blocked blood flow

Scar tissue hinders TIMP1 and 2 in breaking down fibrotic cells leading to reduced blood flow

**Repeat True/False Questions Presented after Reading the Text**

Liver cirrhosis can be cured if caught very early on. [[false]]

Liver cirrhosis can be caused by alcoholism or hepatitis. [[true]]

The liver is an essential organ because it generates red blood cells. [[false]]

People with liver cirrhosis can have both scars and fibrous tissue in the liver. [true]

A visual inspection of the liver using laparoscopy is commonly used for diagnosis.  [[false]]

**Text: Pemphigus**

**True/False Questions before Reading the Text**

An auto-immune disorder is discovered by blisters on the skin. [[false]]

In pemphigus vulgaris, the sores often develop on the inside of the mouth. [[true]]

Pemphigus is triggered by repeated over-exposure to the sun. [[false]]

Pemphigus can be treated. [[true]]

Pemphigus vulgaris is a general term for a group of diseases such as pemphigoid, pemphigitis, etc. [[false]]

**Original Text**

Pemphigus is a general term for a group of rare autoimmune blistering skin disorders. Autoimmune disorders occur when the body's own immune system mistakenly attacks healthy tissue. The symptoms and severity associated with the various forms of pemphigus vary. All forms of pemphigus are characterized by the development of blistering eruptions on the outer layer of the skin (epidermis). In pemphigus vulgaris, lesions also develop on the mucous membranes such as those lining the inside the mouth. Mucous membranes are the thin, moist coverings of many of the body's internal surfaces. If left untreated, pemphigus will usually be fatal. The exact cause of pemphigus is unknown.

The term pemphigus is a general term for a group of related autoimmune blistering skin diseases. The two main types of pemphigus are pemphigus vulgaris and pemphigus foliaceus. Each type has subtypes. Additional disorders are sometimes classified as pemphigus including paraneoplastic pemphigus and pemphigus IgA. Some physicians consider these disorders similar, yet distinct, autoimmune blistering disorders with different causes and clinical, immunological and microscopic tissue (histological) features. Pemphigoid is a general term for a different group of skin disorders. These other disorders are discussed in the related disorders section of this report.

**Simplified Text**

Pemphigus is a group of rare autoimmune blistering skin conditions. Autoimmune conditions come about when the body's own immune system mistakenly attacks healthy tissue. The many kinds of pemphigus have different symptoms and intensity. All kinds of pemphigus get blisters on the top layer of the skin (epidermis). In pemphigus vulgaris, blisters also develop on the mucus tissue layer like those lining the inside the mouth. Mucus tissue layers are the thin, wet natural coverings of many of the body's inside surfaces. If not treated, patients with pemphigus will usually die. The exact cause of pemphigus is not known.

The two main types of pemphigus are pemphigus vulgaris and pemphigus foliaceus. Each type has subtypes. Additional conditions that are included by some doctors as pemphigus include paraneoplastic pemphigus and pemphigus IgA. These conditions also are autoimmune blistering conditions, though they have different causes and clinical, immunological, and microscopic tissue (histological) features. Pemphigoid is a general term for a different group of skin conditions. These other conditions are talked about in the related conditions section of this report.

**Multiple-choice Questions Presented with the Text**

(Perceived Difficulty) After reading this text, I consider this information

Very difficult to understand

Difficult to understand

Easy to understand

Very easy to understand

(Overview Question) This text is about autoimmune disorders:

No, it is about blisters.

No, it is about membranes

Yes, it is about autoimmune disorders resulting in blisters. [[correct]]

Yes, it is about autoimmune disorders resulting in membrane mucus.

(General Question) There are two main types of pemphigus

Yes, pemphigus vulgaris and pemphigus foliaceus [[correct]]

Yes,  pemphigus vulgaris and pemphigoid

Yes, pemphigus foliaceus and pemphigoid

No, there are three main types:  pemphigus vulgaris, pemphigus foliaceus and pemphigoid

**Multiple-choice Questions Presented after Reading the Text**

With pemphigus vulgaris, blisters will develop:

On the outside of the body

On the inside of the body

Both outside and inside of the body. [Answer]

The text does not contain enough information to answer the question.

Pemphigus is

a reaction to lesions in the skin

a condition where we cannot define the cause [Answer]

a reaction to microscopic tissue features

has a consistent set of symptoms

Pemphigus is a condition

where the body tries to damage healthy tissue because it thinks something is wrong with it. [answer]

where the body tries to replace healthy tissue because it thinks something is wrong with it.

where the body tries to protect itself against tissue not from the person's own body

where the body tries to protect itself against tissue that has blisters

The different types of pemphigus are well established.

Yes, the different types of blistering allows for differentiation

Yes, the causes are clear and allow for differentiation

No, some related disorders are sometimes classified as pemphigus [answer]

No, some unknown blistering conditions are sometimes classified as pemphigus

**Repeat True/False Questions Presented after Reading the Text**

An auto-immune disorder is discovered by blisters on the skin. [[false]]

In pemphigus vulgaris the sores often develop on the inside of the mouth. [[ true]]

Pemphigus is usually caused by repeated over-exposure to the sun. [[false]]

Pemphigus can be treated. [[true]]

Pemphigus vulgaris is a general term for a group of diseases such as pemphigoid, pemphigitis, etc. [[false]]

**Text: Polycythemia Vera**

**True/False Questions before Reading the Text**

Polycythemia vera does not result in an increase in white blood cells. [[false]]

Most of the people affected by polycythemia have a genetic mutation.[[true]]

Hyperviscosity is a condition where a person's blood becomes too thin. [[false]]

Polycythemia causes a person to produce too many red blood cells. [[true]]

Polycythemia has only a single symptom. [[false]]

**Original Text**

Polycythemia vera is a rare, chronic disorder involving the overproduction of blood cells in the bone marrow (myeloproliferation). The overproduction of red blood cells is most dramatic. But the production of white blood cells and platelets are also elevated in most cases. Since red blood cells are overproduced in the marrow, this leads to abnormally high numbers of circulating red blood cells (red blood mass) within the blood. Consequently, the blood thickens and increases in volume, a condition called hyperviscosity. Thickened blood may not flow through smaller blood vessels properly. A variety of symptoms can occur in individuals with polycythemia vera including nonspecific symptoms such as headaches, fatigue, weakness, dizziness or itchy skin; an enlarged spleen (splenomegaly); a variety of gastrointestinal issues; and the risk of blood clot formation, which may prevent blood flow to vital organs. More than 90 percent of individuals with polycythemia vera have a mutation of the JAK2 gene. The exact role of this mutation in the development of polycythemia vera is not yet known.

Polycythemia vera belongs to a group of diseases known as the myeloproliferative disorders (MPDs). Three other disorders are commonly classified as MPDs: chronic myeloid leukemia, essential thrombocythemia and idiopathic myelofibrosis.

**Simplified Text**

Polycythemia vera is a rare, chronic condition where the bone marrow makes too many blood cells (myeloproliferation). Red blood cells increase the most, but white blood cells and platelets are also high in most cases. The increase in red blood cells in the bone marrow results in too many red blood cells (red blood mass) in the blood. This causes the blood to thicken and increase in volume causing the blood to be too thick to flow through smaller blood vessels, a condition called hyperviscosity. Someone with polycythemia vera may have many different symptoms, including symptoms that are not specific such as headaches, lack of energy, loss of strength, spinning sensation, or itchy skin; an enlarged spleen (splenomegaly); stomach and intestinal issues; and the risk of forming blood clots, which may prevent blood flow to critical organs. More than 90 percent of people with polycythemia vera have a mutation of the JAK2 gene. The exact function of this mutation in the development of polycythemia vera is not yet known.

Polycythemia vera belongs to a group of conditions known as the myeloproliferative disorders (MPDs). Three other disorders are commonly classified as MPDs: chronic myeloid leukemia, essential thrombocythemia and idiopathic myelofibrosis.

**Multiple-choice Questions Presented with the Text**

(Perceived Difficulty) After reading this text, I consider this information

Very difficult to understand

Difficult to understand

Easy to understand

Very easy to understand

(Overview Question) This text is about a disorder affecting bone marrow

No, it is about a blood disorder.

No, it is about rare cancer

Yes, it is about disorders affecting blood cells in bone marrow. [answer]

Yes, it is about disorders affecting nerve tissue in bone marrow.

(General Question) Polycythemia vera is

One of four disorders in the group of myeloproliferative disorders [answer]

The main disorder with 3 subcategories in the group of myeloproliferative disorders

A disorder often misclassified in the group of myeloproliferative disorders

A disorder that is a subcategory of idiopathic myelofibrosis

**Multiple-choice Questions Presented after Reading the Text**

Polycythemia vera is a blood disease where too many white blood cells are around which makes the bone marrow produce too many red blood cells in response.

too many red blood cells are around leaving not enough room for white blood cells and platelets.

too many red blood cells are around and this makes the blood too thick [answer]

too many red and white blood cells and platelets are around and they start working against each other.

People with polycythemia vera make

more red and white blood cells equally

more red and white blood cells with even more red blood cells [answer]

less red and white blood cells equally

more bone marrow

A known symptom of the disorder is that

the blood may form clots that can stop it from reaching essential organs [answer]

the arteries start narrowing so that the blood cannot get to the heart

the JAK2 gene malfunctions

too many red blood cells start canceling the effect of white blood cells

A person with polycythemia vera can suffer from many different symptoms

that are all fairly specific to the disease, such as blood clots

that are frequently found with other diseases, such as headaches

that are a mix of those easily found with other diseases, such as stomach pains and tiredness [answer]

that can be easily recognized by checking the narrowing of the veins

**Repeat True/False Questions Presented after Reading the Text**

Polycythemia vera does not result in an increase in white blood cells. [[false]]

Most of the people affected by polycythemia have a genetic mutation. [[true]]

Hyperviscosity is a condition where a person's blood becomes too thin. [[false]]

Polycythemia causes a person to produce too many red blood cells. [[true]]

Polycythemia has only a single symptom. [[false]]

## About the Authors

**Prosanta Barai** is a third-year Ph.D. student in Management Information Systems at the University of Arizona's Eller College of Management. His research is dedicated to leveraging cutting-edge AI technologies, particularly large language models and ML, to bridge mental and physical healthcare gaps and enhance healthcare literacy. His work emphasizes developing transparent, interpretable, and ethical AI systems significantly impacting healthcare delivery and accessibility. He holds Applied Statistics and Data Science degrees and has robust foundational skills in machine learning and computational modelling. He is an active member of the Deep Target NLP Lab, contributing to NIH-funded projects aimed at autism symptom identification through NLP and enhancing the quality and reliability of AI-driven healthcare solutions.

**Gondy Leroy** (ORCID 0000-0003-4751-6680) is Professor of MIS, Research Director of the Center for Management Innovations in Healthcare, and Associate Dean for Research at the University of Arizona's Eller College of Management. Her research focuses on machine learning (ML) and artificial intelligence (AI) and has been funded by NIH, AHRQ, NSF, Microsoft Research, and several foundations. She earned a combined BS-MS (1996) in experimental psychology from the Catholic University of Leuven (1996) in Belgium and a MS (1999) and PhD (2003) in management information systems from the University of Arizona. She serves on several journal editorial boards, co-chairs special issues, sessions, tracks, workshops, and conferences, and serves on the NIH Clinical Data Management and Analysis Study Section. She is the author of the book "Designing User Studies in Informatics (Springer, 2011). Finally, and importantly, she mentors students at all educational levels from high school through graduate school.

**Arif Ahmed** is a 4th-year doctoral student in Management Information Systems at the University of Arizona, with a research focus on healthcare analytics and natural language processing. Currently working on an NIH-funded project, his research aims to improve health literacy by leveraging AI to simplify complex medical texts and enhance patient comprehension. His work bridges the gap between data science and public health, exploring how GenAI tools can be responsibly integrated into healthcare communication. Arif brings a strong technical background in machine learning, statistical analysis, and programming, with hands-on experience using Python and R for health data analytics. Passionate about responsible AI use, my long-term goal is to advance patient-centered technologies and pursue a career in academia focused on impactful, translational research in healthcare informatics.